






\magnification\magstephalf
\baselineskip14pt
\vsize23.5truecm 


\def\hatt{\widehat}
\def\dell{\partial}
\def\tilda{\widetilde}

\def\half{\hbox{$1\over2$}}

\def\arr{\rightarrow}
\def\normal{{\cal N}}

\def\RR{\mathord{I\kern-.3em R}}
\def\PP{\mathord{I\kern-.3em P}}
\def\NN{\mathord{I\kern-.3em N}}
\def\ZZ{\mathord{I\kern-.3em Z}} 
\def\Var{{\rm Var}}
\def\E{{\rm E}}
\def\d{{\rm d}}

\def\mtrix{\pmatrix} 

\def\subsection{\medskip}

\font\bigbf=cmbx12

\font\csc=cmcsc10

 at 10truept 
\font\smallrm=cmr8

\def\today{\number\day \space \ifcase\month\or
January\or February\or March\or April\or May\or June\or 
July\or August\or September\or October\or November\or December\fi  
\space \number\year}


   
\def\ref#1{{\noindent\hangafter=1\hangindent=20pt
  #1\smallskip}}          

\def\quotationone{\smallrm Where there is a Will}
\def\quotationtwo{\smallrm There is a Won't}
\def\hskipdistanceleft{\hskip-3.5pt}
\def\hskipdistanceright{\hskip-2.0pt}
\footline={{
\ifodd\count0
        {\hskipdistanceleft\quotationone\phantom{\smallrm\today}
                \hfil{\rm\the\pageno}\hfil
         \phantom{\quotationone}{\smallrm\today}\hskipdistanceright}
        \else 
        {\hskipdistanceleft\quotationtwo\phantom{\today}
                \hfil{\rm\the\pageno}\hfil
         \phantom{\quotationtwo}{\smallrm\today}\hskipdistanceright}
        \fi}}

         
\def\cstok#1{\leavevmode\thinspace\hbox{\vrule\vtop{\vbox{\hrule\kern1pt
        \hbox{\vphantom{\tt/}\thinspace{\tt#1}\thinspace}}
        \kern1pt\hrule}\vrule}\thinspace} 
 


\def\fermat#1{\setbox0=\vtop{\hsize4.00pc
        \smallrm\raggedright\noindent\baselineskip9pt
        \rightskip=0.5pc plus 1.5pc #1}\leavevmode
        \vadjust{\dimen0=\dp0
        \kern-\ht0\hbox{\kern-4.00pc\box0}\kern-\dimen0}}

\def\hsizeplusepsilon{14.25truecm} 
\def\fermatright#1{\setbox0=\vtop{\hsize4.00pc
        \smallrm\raggedright\noindent\baselineskip9pt
        \rightskip=0.5pc plus 1.5pc #1}\leavevmode
        \vadjust{\dimen0=\dp0
        \kern-\ht0\hbox{\kern\hsizeplusepsilon\box0}\kern-\dimen0}}



\def\quotationone{\smallrm{Nils Lid Hjort}}
\def\quotationtwo{\smallrm{Dynamic Likelihood}} 
\def\today{\smallrm{April 1993}}
\def\oneover24{\hbox{$1\over24$}}

\centerline{\bigbf Dynamic likelihood hazard rate estimation}

\smallskip
\centerline{\bf Nils Lid Hjort}

\smallskip
\centerline{\sl University of Oslo and University of Oxford}

{{\smallskip\narrower\noindent\baselineskip11pt
{\csc Abstract.}
The best known methods for estimating hazard rate functions
in survival analysis models 
are either purely parametric or purely nonparametric. 
The parametric ones are sometimes too biased 
while the nonparametric ones are sometimes too variable. 
In the present paper a 
certain semiparametric approach to hazard rate estimation,
proposed in Hjort (1991), is developed further,
aiming to combine parametric and nonparametric features. 
It uses a dynamic local likelihood approach to
fit the locally most suitable member in a given
parametric class of hazard rates,
and amounts to a version of nonparametric parameter smoothing
within the parametric class. 
Thus the parametric hazard rate estimate at time $s$ inserts a 
parameter estimate that also depends on $s$. 
We study bias and variance properties of the resulting 
estimator and methods for choosing the local smoothing parameter. 
It is shown that dynamic likelihood estimation often leads to 
better performance than the purely nonparametric methods, 
while also having capacity for not losing much to the parametric 
methods in cases where the model being smoothed is adequate.

\smallskip\noindent
{\csc Key words:} 
{\sl
dynamic likelihood, 
hazard rate, 
kernel smoothing, 
local goodness of fit, 
local modelling, 
semiparametric estimation}
\smallskip}}

\bigskip
{\bf 1. Introduction and summary.} 
This paper concerns a class of semiparametric type methods of 
estimating hazard rate functions in models for life history data. 
The best known methods for estimating such hazard rates 
are those that are either purely parametric or purely nonparametric.
The parametric methods are usually biased 
since parametric models are usually imperfect,
and the nonparametric methods often have high estimation variance. 
There should accordingly be room for methods that 
somehow lie between the parametric and the nonparametric ones. 
One might hope that such methods 
are better than the nonparametric ones if the 
true hazard is in the vicinity of the parametric model,
while not being much worse than the parametric ones if the 
parametric model is true.

Although results can be obtained in a more general framework
of counting process models we shall mainly be content 
to illustrate and investigate ideas for the 
`random censorship' model, 
which is the simplest and perhaps most important special case 
of such models for censored life-time data. 
It postulates that life-times 
$X_1^0,\ldots,X_n^0$ from a population are i.i.d.~with
density $f(.)$, cumulative distribution $F(.)$, 
and hazard rate function $\alpha(.)$ given by 
$\alpha(s)=f(s)/F[s,\infty)$; $\alpha(s)\,{\d s}$ is the
probability of failing in $[s,s+{\d s})$ given that an individual
is still at risk at time $s$. 
The life-time $X_i^0$ may not be directly observed, however,
because of a possibly interfering censoring variable $C_i$;
only $X_i={\rm min}(X_i^0,C_i)$ and the indicator variable
$\delta_i=I\{X_i^0\le C_i\}$ are observed. 
For simplicity and concreteness we stipulate that the $C_i$'s 
are independent of the life-times and 
i.i.d.~according to a distribution with cumulative function $G$. 
In particular the $n$ pairs $(X_i,\delta_i)$ are i.i.d. 
Finally we shall assume that data are obtained on a finite time horizon 
basis, say on $[0,T]$ for a known and finite $T$. 
This is convenient for some of the martingale convergence theory
and is not a practical limitation. 

The parametric approach is to postulate that 
$\alpha(s)=\alpha(s,\theta)$ for a suitable family,
indexed by some one- or multi-dimensional $\theta$. 
Typical examples include 
the exponential, 
the Weibull, 
the simple frailty model 
with $\alpha(s)=\theta_1/(1+\theta_2s)$, 
the piecewise constant hazard rate model,
the Gompertz--Makeham distribution, 
the gamma, and the log-normal. 
Properties of the maximum likelihood method for estimating 
$\theta$ with censored data have been 
studied by Borgan (1984) and others under the condition 
that the model is correct,
i.e.~that there really is some $\theta_0$ with
$\alpha(s)=\alpha(s,\theta_0)$ on $[0,T]$. 
In practice the model is never perfect, however, and it is
useful to study estimation methods outside model conditions, 
where the best parameter is to be thought of as being 
`least false' or `most suitable', as opposed to `true'.  
The large-sample behaviour of several estimation methods
in this wider setting has been explored in Hjort (1992). 
Some results about this are reviewed in Section 2 and are 
used in later sections.  

In Section 3 a dynamic likelihood approach to parametric 
hazard rate estimation is presented. It takes as its basis
any given parametric hazard function and consists of 
inserting a local parameter estimate $\hatt\theta(s)$ 
in $\alpha(s,\theta)$ at time $s$, producing 
$$\hatt\alpha(s)=\alpha(s,\hatt\theta(s)), $$
where the parameter estimate is obtained using only information on those 
individuals that have survived up to $s-\half h$ and 
what happens to them on $[s-\half h,s+\half h]$. 
This amounts to a kind of nonparametric parameter smoothing 
within a given parametric class. A more general estimator
involving smoothing with a kernel function is also discussed. 
Bias and variance properties are studied in Section 3 for
one-dimensional and in Section 4 for multi-dimensional families.
It turns out that 
$$\E\hatt\alpha(s)\doteq\alpha(s)+\half\beta_Kh^2b(s)
\quad {\rm and} \quad
\Var\,\hatt\alpha(s)\doteq{\gamma_K\over nh}{\alpha(s)\over y(s)}, $$
where $\beta_K$ and $\gamma_K$ are characteristics of the kernel
function used and $y(s)$ is the limiting proportion of individuals
still at risk at time $s$. The $b(s)$ is a certain bias factor,
the size of which depends on both $\alpha''(s)$ and characteristics
of the underlying parametric model used. These results match 
closely those of the most usual nonparametric method,
that of smoothing the empirical cumulative hazard function, for which
$$\E\tilda\alpha(s)\doteq\alpha(s)+\half\beta_Kh^2\alpha''(s)
\quad {\rm and} \quad
\Var\,\tilda\alpha(s)\doteq{\gamma_K\over nh}{\alpha(s)\over y(s)}. $$
In Section 5 situations are characterised where the new method 
performs better than the traditional nonparametric method. 
Methods for choosing the local smoothing parameter $h$ are 
discussed in Section 6, including the arduous one that 
for each $s$ expands the $s\pm\half h$ interval until a goodness of fit
criterion rejects the model. Overall it transpires that a suitable
dynamic likelihood estimator often can perform better than the 
purely nonparametric ones, while at the same time not losing much 
to parametric ones when the true hazard is close to the parametric hazard. 
Finally some supplementing results and remarks are offered in Section 7. 

This paper expands in several ways on the basic results 
that were already presented in Hjort (1991). That paper also 
proposed two further semiparametric estimation schemes,
one using orthogonal expansions to correct on an initial
parametric guess, and one Bayesian procedure that employs a
nonparametric prior around a given parametric hazard model.  

\bigskip
{\bf 2. Purely nonparametric and purely parametric estimation.}
This section introduces some basic notation and 
reviews properties of the Nelson--Aalen estimator 
for the cumulative hazard function in the nonparametric case 
and of the maximum likelihood and maximum weighted likelihood 
estimators in the parametric case. These will be used in later sections. 
Since our ambition is to go beyond ordinary parametric methods 
the behaviour of these must be considered also outside the conditions 
of the postulated parametric model. 

\subsection 
{\csc 2.1. Nonparametric estimation.}
Let $N(t)=\sum_{i=1}^nI\{X_i\le t,\delta_i=1\}$ 
be the counting process and 
$Y(s)=\sum_{i=1}^nI\{X_i\ge s\}$ the at risk process, 
and form from these the Nelson--Aalen estimator 
$$\hatt A(t)=\int_0^t{{\d N}(s)\over Y(s)}
	=\sum_{i=1}^n{\delta_i\over Y(X_i)}I\{X_i\le t\} \eqno(2.1)$$
for the cumulative hazard rate $A(t)=\int_0^t\alpha(s)\,{\d s}$. 
Its properties are best explained using the martingale 
$B(t)=N(t)-\int_0^tY(s)\alpha(s)\,{\d s}$. 
Let $y(s)$ be the limit in probability of $\hatt y(s)=Y(s)/n$, 
i.e.~the limiting proportion of individuals under risk at time $s$,
and equal to $F[s,\infty)G[s,\infty)$  
under present circumstances, where $G(.)$ is the censoring distribution.
A basic large-sample property of $B$ is that $B(t)/\sqrt{n}$ 
goes to a Gau\ss ian martingale $V(t)$ with independent increments 
and noise level $\Var\,{\d V}(s)=y(s)\alpha(s)\,{\d s}$, 
and, more generally, 
that $\int_0^tH_n(s)\,{\d B}(s)/\sqrt{n}$ tends to $\int_0^th(s)\,{\d V}(s)$
in distribution, in cases where $H_n(.)$ is previsible 
(its value at $s$ is known at $s-$) 
and converges to the deterministic $h(.)$. 
It follows from these facts that 
$$\eqalign{
\sqrt{n}\{\d\hatt A(s)-{\d A}(s)\}
&={I\{Y(s)\ge 1\}\over \hatt y(s)}
	\,{{\d B}(s)\over \sqrt{n}}-I\{Y(s)=0\}\,{\d A}(s) \cr	
&\doteq_d \hatt y(s)^{-1}{\d B}(s)/\sqrt{n}
	\rightarrow_d y(s)^{-1}\,{\d V}(s) \cr}\eqno(2.2)$$ 
in the large-sample limit. 
In particular $\d\hatt A(s)$ is very nearly unbiased for ${\d A}(s)$
and $\sqrt{n}\{\hatt A(t)-A(t)\}$ tends to the Gau\ss ian martingale
$\int_0^ty(s)^{-1}\,{\d V}(s)$ with variance
$\int_0^ty(s)^{-1}\alpha(s)\,{\d s}$. 
See for example the recent book Andersen, Borgan, Gill \& Keiding 
(1993, Chapter II) for more details.
The usual nonparametric way of estimating the hazard rate itself is
to smooth the Nelson--Aalen and take the derivative, see (5.1).

\subsection
{\csc 2.2. Maximum likelihood estimation.} 
A parametric model is of the form $\alpha(t)=\alpha(t,\theta)$,
where $\theta=(\theta_1,\ldots,\theta_p)'$ is some 
$p$-dimensional parameter. 
The log-likelihood for the observed data can be written 
$L_n(\theta)=\int_0^T\{\log\alpha(t,\theta)\,{\d N}(t)
	-Y(t)\alpha(t,\theta)\,{\d t}\}$,
see for example Andersen et al.~(1993, Chapter VI). 
This defines the maximum likelihood estimator $\hatt\theta$. 

To explain the large-sample behaviour of this estimator, let 
$U_n(\theta)=n^{-1}\int_0^T\psi(t,\theta)\{{\d N}(t)
	-Y(t)\alpha(t,\theta)\}\,{\d t}$ 
be the $p$-vector of first partial derivatives
of $n^{-1}L_n(\theta)$, where we write 
$\psi(t,\theta)={\dell\over \dell\theta}\log\alpha(t,\theta)$. 
Under natural regularity conditions $U_n(\theta)$ 
tends in probability to 
$u(\theta)=\int_0^Ty(t)\allowbreak\psi(t,\theta)\allowbreak
	\{\alpha(t)-\alpha(t,\theta)\}\,{\d t}$,  
with $y(t)$ as above. The maximum likelihood estimator, 
which solves $U_n(\hatt\theta)=0$, converges in probability 
to the particular parameter value $\theta_0$ that solves $u(\theta_0)=0$. 
We think of this as the `least false' or `agnostic' parameter value, 
and it minimises the distance measure 
$$d[\alpha,\alpha(.,\theta)]=\int_0^T y\bigl[
	\alpha\{\log\alpha-\log\alpha(.,\theta)\}
		-\{\alpha-\alpha(.,\theta)\}\bigr]\,{\d t} \eqno(2.3)$$
between true model and approximating model. 
This is proved in Hjort (1992). 
In later sections we shall also need the large-sample distribution, 
and quote the following result from Hjort (1992). 
Consider the $p\times p$-matrix
$\psi^*(t,\theta)=\dell^2\log\alpha(t,\theta)/\dell\theta\dell\theta'$
and the function 
$E(t)=\int_0^t y(s)\psi(s,\theta_0)\{\alpha(s)-\alpha(s,\theta_0)\}\,{\d s}$
(in particular $E(0)=E(T)=0$). 
Define $p\times p$-matrices 
$$\eqalign{
J&=\int_0^T\Bigl[y(t)\psi(t,\theta_0)\psi(t,\theta_0)'\alpha(t,\theta_0)
-y(t)\psi^*(t,\theta_0)\{\alpha(t)-\alpha(t,\theta_0)\}\Bigr]\,{\d t}, \cr
M&=\int_0^T\Bigl[y(t)\psi(t,\theta_0)\psi(t,\theta_0)'\alpha(t)
	+\bigl\{\psi(t,\theta_0)E(t)'+E(t)\psi(t,\theta_0)'\bigr\}
		\alpha(t,\theta_0)\Bigr]\,{\d t}. \cr}$$	
Then $\sqrt{n}(\hatt\theta-\theta_0)\arr_d\normal_p\{0,J^{-1}MJ^{-1}\}$. 
Note that under model conditions $\alpha(t)$ is indeed equal to
$\alpha(t,\theta_0)$,
the expressions for $J$ and $M$ simplify and become equal, 
and we have the more familiar-looking
limit distribution $\normal_p\{0,J^{-1}\}$, a result proved by Borgan (1984). 

\subsection
{\csc 2.3. M-estimators.} 
We shall also need some general results 
about weighted likelihood estimators, from Hjort (1992, Section 5).
Consider $\int_0^T G_n(t)\{\log\alpha(t,\theta)\,{\d N}(t)\allowbreak
	-Y(t)\alpha(t,\theta)\,{\d t}\}$ 
instead of the ordinary log-likelihood (which uses $G_n(t)=1$),
and let $\hatt\theta_{g}$ maximise. 
This estimator also solves 
$\int_0^TG_n(t)\psi(t,\theta)\{{\d N}(t)-Y(t)\alpha(t,\theta)\,{\d t}\}=0$,
and belongs to the class of M-estimators for this counting process
model, see Hjort (1985) and Andersen et al.~(1993, Chapter VI). 
Assume that the weight function $G_n(t)$ is previsible and goes in 
probability to $g(t)$. The first result is that this estimator is 
consistent for the particular least false parameter value
$\theta_{0,g}$ that minimises the distance function 
$$d_g[\alpha,\alpha(.,\theta)]=\int_0^Tgy\bigl[
	\alpha\{\log\alpha-\log\alpha(.,\theta)\}
		-\{\alpha-\alpha(.,\theta)\}\bigr]\,{\d t}, \eqno(2.4)$$
a generalisation of (2.3).
It also solves $\int_0^Tg(t)y(t)\psi(t,\theta)\{\alpha(t)
	-\alpha(t,\theta)\}\,{\d t}=0$.
Secondly, 
$$\sqrt{n}(\hatt\theta_{g}-\theta_{0,g})
	\arr_d\normal_p\{0,J_g^{-1}M_gJ_g^{-1}\}, \eqno(2.5)$$
where $J_g$ and $M_g$ are appropriate generalisations of those 
appearing above. In fact
$$\eqalign{
J_g&=\int_0^Tgy\bigl[\psi_0\psi_0'\alpha_0 
	-\psi^*_0(\alpha-\alpha_0)\bigr]\,{\d t}, \cr
M_g&=\int_0^T\bigl[g^2y\psi_0\psi_0'\alpha
	+g\{\psi_0E_g'+E_g\psi_0'\}
		\alpha_0\bigr]\,{\d t}, \cr} \eqno(2.6)$$	
in which $E_g(t)=\int_0^t gy\psi_0(\alpha-\alpha_0)\,{\d s}$,
and where $\alpha_0=\alpha(s,\theta_{0,g})$,
$\psi_0=\psi(s,\theta_{0,g})$. 
Note that both $E_g(0)$ and $E_g(T)$ are equal to 0,
and that the expressions for $J_g$ and $M_g$ simplify 
when the model happens to be correct.  

\bigskip 
{\bf 3. Dynamic likelihood estimation.} 
Of course the parametric estimation method of 2.2 works best
if the postulated model is adequate, i.e.~if there 
really is a single $\theta_0$ that secures 
$\alpha(s)\doteq\alpha(s,\theta_0)$ throughout $[0,T]$.
Otherwise there is modelling bias present and it could for example be 
advantageous to use different $\theta_0$'s 
in different sub-intervals. 
We shall pursue a somewhat more extreme version of this idea, 
namely to fit a local estimate $\hatt\theta(s)$ 
for each $s$, and then use $\alpha(s,\hatt\theta(s))$ in the end.

\subsection 
{\csc 3.1. Dynamic likelihood.} 
The dynamic or local likelihood estimation proposal is to use 
the M-estimator apparatus with a `window function' 
$G_n(t)=g(t)=I\{t\in W\}$, where $W=[s-\half h,s+\half h]$ 
is a local interval around a given fixed $s$. 
So let $\hatt\theta(s)$ maximise 
$$L_W(\theta)=\int_W\{\log\alpha(t,\theta)\,{\d N}(t)
	-Y(t)\alpha(t,\theta)\,{\d t}\}. \eqno(3.1)$$
The resulting {\it dynamic likelihood hazard rate estimator} is 
$$\hatt\alpha(s)=\alpha(s,\hatt\theta(s)). \eqno(3.2)$$
Note that $L_W(\theta)$, the local log-likelihood at window $W$ 
around $s$, is a bona fide log-likelihood, 
namely that based on those individuals that have 
survived up to $s-\half h$ and information about what happens to
them in $[s-\half h,s+\half h]$. Showing this is not difficult
by first noting that this group of individuals have 
$$\matrix{
{\rm probability\ density}\hfill
&=\alpha(t,\theta)\exp\bigl\{-\int_{s-h/2}^t
	\alpha(u,\theta)\,{\d u}\bigr\} \hfill
&{\rm for\ }t\in[s-\half h,s+\half h], \hfill \cr
{\rm and\ chance\ }\hfill
&=\exp\bigl\{-\int_{s-h/2}^{s+h/2}\alpha(u,\theta)\,{\d u}\bigr\} \hfill
&{\rm of\ further\ surviving\ }s+\half h. \hfill \cr}$$
Consciously disregarded, for example, is information about 
individuals failing in $[0,s-\half h)$. Including such a 
$[1-\exp\{-A(s-\half h,\theta)\}]^{n_0}$ term would have 
strengthened the likelihood and made our $\theta$ estimator
more precise -- but only if the parametric form
of the hazard is correct also to the left of $s-\half h$.
The crucial idea here is to only trust the parametric form locally,
and this leads to the (3.1) log-likelihood. 
Of course if $h$ is large, which should correspond to trusting
the model over the full range, then we get back the full 
log-likelihood and ordinary maximum likelihood. 

The $\hatt\theta(s)$ estimator aims 
at the locally most suitable parameter value $\theta_0(W)=\theta_0(s)$ 
that minimises (2.4) with $g=I_W$, or, equivalently, 
solves $\int_Wy(t)\psi(t,\theta)\{\alpha(t)-\alpha(t,\theta)\}\,{\d t}=0$.
Its large-sample behaviour is described by (2.5), which suggests
$$\E\hatt\theta(s)\doteq \theta_0(s),
	\quad 
  {\rm VAR}\,\hatt\theta(s)\doteq J_W^{-1}M_WJ_W^{-1}/n, $$
where $J_W$ and $M_W$ are as in (2.6) with $g(t)=I\{t\in W\}$.
This transforms into corresponding properties for $\hatt\alpha(s)$
by Taylor expansions and delta-method arguments:
$$\eqalign{
\E\alpha(s,\hatt\theta(s))&\doteq \alpha(s,\theta_0(s)), \cr
\Var\,\alpha(s,\hatt\theta(s))
	&\doteq \alpha(s,\theta_0(s))^2\psi(s,\theta_0(s))'
		J_W^{-1}M_WJ_W^{-1}\psi(s,\theta_0(s))/n. \cr} \eqno(3.3) $$

These approximations are valid if $h$ is fixed and $n$ is large.  
But we are also interested in becoming increasingly 
fine-tuned about the $s\pm\half h$ interval as $n$ grows. 
In order to study the bias and variance properties more closely,
observe first that if $z(t)$ is a twice differentiable function 
defined in a neighbourhood of $s$, then
$\int_Wz(t)\,{\d t}\doteq z(s)h+{1\over 24}z''(s)h^3$ 
by a simple Taylor argument. From this and the defining 
equation for $\theta_0(s)$ we see that 
$$y(s)\psi(s,\theta)\{\alpha(s)-\alpha(s,\theta)\}
  +\oneover24\bigl(y\psi(.,\theta)(\alpha-\alpha(s,\theta))\bigr)''(s)h^2
	\doteq0,$$
for the particular value $\theta=\theta_0(s)$, 
where $(fgh)''(s)$ means the second derivative of the $f(s)g(s)h(s)$ 
function evaluated at $s$. This implies generally that 
$\alpha(s,\theta_0(s))=\alpha(s)+O(h^2)$. 
One can also show from this that 
$$\E\alpha(s,\hatt\theta(s))=\alpha(s,\theta_0(s))+O(1/n)
	=\alpha(s)+O(h^2+1/n). $$
In order for the bias of the (3.2) estimator 
to go to zero it is therefore necessary that 
$h\rightarrow0$ as $n\rightarrow\infty$. 

At the moment we shall be content to give a bias formula for the 
case of a one-parameter family $\alpha(s,\theta)$, for which 
$$\alpha(s,\theta_0(s))
	\doteq\alpha(s)+{h^2\over 24}\Bigl[\alpha''(s)-\alpha_0''(s)
	+2\{\alpha'(s)-\alpha_0'(s)\}\Bigl\{{y'(s)\over y(s)}
	+{\psi_0'(s)\over \psi_0(s)}\Bigr\}\Bigr]. \eqno(3.4)$$ 
In this formula $\alpha_0'(s)$ means the derivative of $\alpha(s,\theta)$
w.r.t.~$s$, and then inserted $\theta=\theta_0(s)$, and similarly
for $\alpha_0''(s)$ and $\psi_0'(s)$. The case of multi-parametric
classes of hazard rates is handled in Section 4. 

Turning next to the variance matrix, one finds 
after using the (2.6) expressions and
the previously established $O(h^2)$ result for the bias that 
$J_W=y(s)\psi_0(s)\psi_0(s)'\alpha(s,\theta_0(s))\,h+O(h^3)$ 
and $M_W=y(s)\psi_0(s)\psi_0(s)'\alpha(s)\,h+O(h^3)$,
under smoothness assumptions on $\alpha(.)$ and $y(.)$, 
and writing for simplicity $\psi_0(s)$ for $\psi(s,\theta_0(s))$. 
We note here for the one-parameter case that 
$\Var\,\hatt\theta(s)\doteq(nh)^{-1}\{y(s)\alpha(s)\psi_0(s)^2\}^{-1}$, 
which in its turn implies
$$\Var\,\alpha(s,\hatt\theta(s))
	\doteq {1\over nh}{\alpha(s)\over y(s)}. \eqno(3.5)$$
Thus $nh\rightarrow\infty$ is necessary for the variance to go 
to zero, and this together with $h\rightarrow0$ 
suffices for consistency of the (3.2) estimator. 

\subsection 
{\csc 3.2 Special case: estimating the local constant.}
The simplest model to try out is the one having $\alpha(s,\theta)=\theta$,
a constant hazard. 
The local hazard estimate and its limit in probability are
$$\hatt\alpha(s)=\hatt\theta(s)
	={\int_W{\d N}(t)\over \int_WY(t)\,{\d t}} 
	\rightarrow_p
	 {\int_W y(t)\alpha(t)\,{\d t}\over \int_W y(t)\,{\d t}}
		=\theta_0(s), \eqno(3.6)$$
again with $W=[s-\half h,s+\half h]$. 
The estimate is of the type total occurrence over total exposure,
and the underlying local least false parameter is a 
local $y$-weighted average of the true hazard rate. 
By earlier efforts 
$$\E\hatt\alpha(s)\doteq\alpha(s)+{h^2\over 24}
	\Bigl\{\alpha''(s)+2\alpha'(s){y'(s)\over y(s)}\Bigr\}
	\quad {\rm and} \quad 
\Var\,\hatt\alpha(s)
	\doteq{1\over nh}{\alpha(s)\over y(s)}. \eqno(3.7)$$
This can also be verified directly. 
Further attention to these details is given in the next subsection. 

A general remark about the dynamic likelihood method is that 
the particular parametric model used 
should be allowed to be quite crude, since we only
employ it as a local approximation to the true hazard rate. 
This example
illustrates this. (3.7) shows that even when $\alpha(.)$ 
simplistically is modelled as being locally a constant 
the result is a reasonable nonparametric estimator.
  
\subsection 
{\csc 3.3. Kernel smoothed dynamic likelihood.}
The dynamic likelihood method of Sections 3.1 and 3.2 can be 
generalised to kernel smoothed variants. 
Let $K(u)$ be a symmetric kernel function with support $[-\half,\half]$ 
and integral 1. Define the local kernel smoothed likelihood 
estimator $\hatt\theta(s)$ to maximise 
$$\tilda L_W(\theta)=\int_WK\bigl(h^{-1}(t-s)\bigr)
  \bigl\{\log\alpha(t,\theta)\,{\d N}(t)
  -Y(t)\alpha(t,\theta)\,{\d t}\bigr\}. \eqno(3.8)$$
The hazard rate estimator is as in (3.2) with this more
general estimate of $\theta$. 
The previously defined local likelihood estimate corresponds to
the special case $K(u)=1$ on $[-\half,\half]$. 
This uniform choice has perhaps some special appeal
since the dynamic log-likelihood $L_W(\theta)$ then can be interpreted as 
a genuine log-likelihood for a subgroup of the individuals under study.
The current smoothed likelihood is more of a mathematical
construction, but turns out to produce 
estimators with slightly better properties, for good choices of $K(u)$. 

We can draw on the general results of 2.3 to find approximate
bias and variance for the maximiser of (3.8). 
Let $\beta_K=\int u^2K(u)\,{\d u}$ and $\gamma_K=\int K(u)^2\,{\d u}$. 
(2.6) with Taylor expansion quickly gives 
$$\eqalign{
J_W&=y(s)\psi_0(s)\psi_0(s)'\alpha(s,\theta_0(s))\,h+O(h^3), \cr 
M_W&=\gamma_Ky(s)\psi_0(s)\psi_0(s)'\alpha(s)\,h+O(h^3), \cr} \eqno(3.9)$$
The multi-parameter case requires more precise expansions, 
since the inverse of $J_W$ is needed and $\psi_0(s)\psi_0(s)'$ has rank 1. 
Leaving the multi-parameter case for Section 4, consider 
an arbitrary one-parameter family $\alpha(s,\theta)$, 
where $\hatt\theta(s)$ solves 
$\int_WK(h^{-1}(t-s))\psi(t,\theta)\{{\d N}(t)
-Y(t)\alpha(t,\theta)\,{\d t}\}=0$.
It aims at the locally least false $\theta_0=\theta_0(W)$ that solves 
$\int_WK(h^{-1}(t-s))\psi(t,\theta)y(t)\{\alpha(t)
-\alpha(t,\theta)\}\,{\d t}=0$,
or 
$\int K(u)\psi(s+hu,\theta)y(s+hu)\{\alpha(s+hu)
-\alpha(s+hu,\theta)\}\,{\d u}=0$. 
Taylor expansion shows that 
$\int K(u)z(s+hu)\,{\d u}=z(s)+\half\beta_Kh^2z''(s)+O(h^4)$ 
for smooth $z(.)$ functions, and this, 
in conjunction with (3.3) and (3.9), leads to 
$$\E\alpha(s,\hatt\theta(s))
       \doteq\alpha(s)+\half h^2\beta_Kb(s)
	\quad {\rm and} \quad 
\Var\,\alpha(s,\hatt\theta(s))
       \doteq{\gamma_K\over nh}{\alpha(s)\over y(s)}, \eqno(3.10)$$
where the bias factor is 
$$b(s)=\alpha''(s)-\alpha_0''(s)
	+2\{\alpha'(s)-\alpha_0'(s)\}\Bigl\{{y'(s)\over y(s)}
	+{\psi_0'(s)\over \psi_0(s)}\Bigr\}. \eqno(3.11)$$
The fact that $\int uK(u)\,{\d u}=0$ is used here. When $K(u)$ is uniform 
we get back (3.4) and (3.5). 
Observe that the approximate variance does not depend on the 
parametric family employed (to the order of approximation used). 

\subsection
{\csc 3.4. Special case: local constant with a kernel.} 
Let us illustrate this for the special case where 
$\alpha(s,\theta)=\theta$. Then 
$$\hatt\alpha(s)=\hatt\theta(s)={\int_WK(h^{-1}(t-s))\,{\d N}(t)
	\over \int_WK(h^{-1}(t-s))Y(t)\,{\d t}}
={\sum_{|x_i-s|\le h/2}K(h^{-1}(x_i-s))\delta_i
\over \sum_{i=1}^n\int_{W\cap[0,x_i]}K(h^{-1}(t-s))\,{\d t}}\,, \eqno(3.12)$$
a locally weighted occurrence over locally weighted exposure estimate.
Here and later on $x_i$ denotes the observed value of $X_i=\min(X_i^0,C_i)$. 
Previous efforts give 
$$E\hatt\alpha(s)=\alpha(s)+\half\beta_Kh^2\bigl\{\alpha''(s)
		+2\alpha'(s)y'(s)/y(s)\bigr\}+O(h^4), $$
and variance 
$\gamma_K(nh)^{-1}\alpha(s)/y(s)$ as before. This generalises (3.7).

One theoretical advantage that (3.12) has over the (3.6) estimator 
is that it has smaller mean squared error, for several natural
choices of kernel $K$, see 6.1. A more immediate practical advantage is 
that $K$ can be chosen to make it smoother than the (3.6) version,
which is discontinuous at time points $s$ where $s\pm\half h$ 
is equal to observed failure times. (3.12) is continuous when 
$K(\pm\half)=0$, and has a continuous derivative if $K$ is chosen 
such that $K'(\pm\half)=0$. 

\bigskip
{\bf 4. Dynamic likelihood for multi-parameter families.} 
The dynamic likelihood and kernel smoothed dynamic likelihood ideas 
of Section 3 can be applied for any smooth parametric family
of hazards, but the basic bias and variance properties have 
so far only been derived for one-parameter families. 
We saw in (3.9), for example, that the multi-parameter case
requires more careful expansions. It is not clear at the outset 
that we gain in precision by smoothing e.g.~a two-parameter hazard
family. We should perhaps expect larger windows to be 
required to be able to estimate both parameters with reasonable precision. 

\subsection
{\csc 4.1. A running Gompertz estimator.}
The hazard function model $\alpha(t)=a\exp(\beta t)$
is sometimes called the Gompertz model. 
Concentrating on a fixed $s$ with fixed window $W=s\pm\half h$, we may 
reparametrise the hazard as 
$$\alpha(t,\theta,\beta)=a\exp(\beta s)\exp(\beta(t-s))
	=\theta\exp(\beta(t-s))
	\quad {\rm for\ }t\in[s-\half h,s+\half h], \eqno(4.1)$$
and interpret $\theta$ as the `local level' and $\beta$ as
the `local slope'. 
Define $\hatt\theta(s)$ and $\hatt\beta(s)$ as those maximising 
the kernel smoothed dynamic likelihood 
$$\tilda L_W(\theta,\beta)
=\int_WK(h^{-1}(t-s))\bigl[\{\log\theta+\beta(t-s)\}\,{\d N}(t)
	-Y(t)\theta\exp(\beta(t-s))\,{\d t}\bigr]. \eqno(4.2)$$
One has
$$\hatt\theta(s,\beta)={\int_WK(h^{-1}(t-s))\,{\d N}(t)
\over \int_WK(h^{-1}(t-s))Y(t)\exp(\beta(t-s))\,{\d t}}, \eqno(4.3)$$
and the resulting profile dynamic likelihood can be shown to 
be concave in $\beta$, and accordingly not very difficult to maximise.
The maximiser found is then inserted into (4.3) to give 
$\hatt\theta(s)$. Note that the general dynamic likelihood recipe gives
$$\hatt\alpha(s)=\alpha(s,\hatt\theta(s),\hatt\beta(s))
	=\hatt\theta(s), \eqno(4.4)$$
simply, so the $\beta$ parameter estimate is only somewhat 
silently present. 			
From the general theory of Section 2.3 we know that 
$\hatt\theta(s)$ and $\hatt\beta(s)$ aim at certain appropriate 
least false parameter values $\theta_0=\theta_0(s)$ and
$\beta_0=\beta_0(s)$, depending on the window $W$, 
and that $\sqrt{n}(\hatt\theta(s)-\theta_0,\hatt\beta(s)-\beta_0)$
goes to a zero-mean normal with covariance matrix 
$J_W^{-1}M_WJ_W^{-1}$. 
Here $J_W$ and $M_W$ are as in (2.6)
with $g=I_W$. We now set out to provide informative approximations 
for these and for the least false local parameters. 

The least false parameter values are such that they solve 
the two equations 
$\int_WK(h^{-1}(t-s))\psi(t,\theta_0,\beta_0)y(t)\{\alpha(t)
-\theta_0\exp(\beta_0(t-s))\}\,{\d s}=0$, where $\psi(t,\theta,\beta)
=(1/\theta,t-s)$. The first equation gives 
$$\theta_0={\int_WK(h^{-1}(t-s))y(t)\alpha(t)\,{\d t}
\over \int_WK(h^{-1}(t-s))y(t)\exp(\beta_0(t-s))\,{\d t}}
={\int K(u)y(s+hu)\alpha(s+hu)\,{\d u}
\over \int K(u)y(s+hu)\exp(\beta_0hu)\,{\d u}}, $$
where the latter integrals are over the support $[-\half,\half]$ for
the kernel function $K(u)$. Upon using 
$\int K(u)z(s+hu)\,{\d u}=z(s)+\half\beta_Kh^2z''(s)+O(h^4)$ again, 
one finds after some calculations that 
$$\theta_0\doteq\alpha(s)+\half\beta_Kh^2[\{y(t)\alpha(t)\}''(s)
-\alpha(s)\{y(t)\exp(\beta_0(t-s))\}''(s)]/y(s)
	=\alpha(s)+\half\beta_Kh^2b(s,\beta_0), $$
say, up to $O(h^4)$ terms, where in fact 
$b(s,\beta_0)=\alpha''(s)-\alpha(s)\beta_0^2+2\{y'(s)/y(s)\}\{\alpha'(s)
-\alpha(s)\beta_0\}$. Similarly the second equation gives
$\int K(u)u\,y(s+hu)\{\alpha(s+hu)-\theta_0\exp(\beta_0hu)\}\,{\d u}=0$,
which upon using $\int K(u)u\,z(s+hu)\,{\d u}=\beta_Kz'(s)h+O(h^3)$ delivers
$\beta_0=\alpha'(s)/\alpha(s)+O(h^2)$. 
This can be plugged into $b(s,\beta_0)$ above to give 
$$\alpha(s,\theta_0(s),\beta_0(s))=\theta_0
=\alpha(s)+\half\beta_Kh^2\{\alpha''(s)-\alpha'(s)^2/\alpha(s)\}
	+O(h^4). \eqno(4.5)$$
Note that the bias is only $O(h^4)$ at $s$ 
if the true $\alpha(.)$ is locally like a Gompertz hazard.

Next consider the matrices that determine the approximate 
variances for $\hatt\theta(s)$ and $\hatt\beta(s)$. From (2.6),
$$\eqalign{
J_W&=\int_WK(h^{-1}(t-s))y(t)\Bigl[\mtrix{1/\theta_0^2 & (t-s)/\theta_0 \cr
(t-s)/\theta_0 & (t-s)^2 \cr}\theta_0\exp(\beta_0(t-s)) \cr 
&\qquad +\mtrix{1/\theta_0^2 & 0 \cr 0 & 0 \cr}
	\{\alpha(t)-\theta_0\exp(\beta_0(t-s))\}\Bigr]\,{\d t}. \cr}$$
We find
$$J_{11}=h\int K(u)y(s+hu)\theta_0^{-2}\alpha(s+hu)\,{\d u}
=hy(s)\alpha(s)/\theta_0^2+\half\beta_Kh^3(y\alpha)''(s)/\theta_0^2+O(h^5)$$
for the (1,1) element. Similar calculations give 
$$J_W=h\Bigl[\mtrix{y(s)\alpha(s)/\theta_0^2 & 0 \cr 0 & 0 \cr}
+\beta_Kh^2\mtrix{a_{11} & a_{12} \cr a_{12} & a_{22} \cr}\Bigr]+O(h^5),$$
where in fact $a_{11}=\half(y\alpha)''(s)/\theta_0^2$,
$a_{12}=y'(s)+y(s)\beta_0$, and $a_{22}=y(s)\theta_0$. 
Next look at 	
$$\eqalign{
M_W&=\int_W\Bigl[K(h^{-1}(t-s))^2y(t)\mtrix{1/\theta_0^2 & (t-s)/\theta_0 \cr
(t-s)/\theta_0 & (t-s)^2 \cr}\alpha(t) \cr 
&\qquad +K(h^{-1}(t-s))
\mtrix{2E_1(t)/\theta_0 & E_2(t)/\theta_0+E_1(t)(t-s) \cr
E_2(t)/\theta_0+E_1(t)(t-s) & 2(t-s)E_2(t) \cr}\Bigr]\,{\d t}, \cr}$$
where $E_1(t)$ and $E_2(t)$ are the components of the $E(t)$ function
defined after (2.6). It turns out that $E_1(t)=O(h^3)$ while
$E_2(t)=O(h^4)$, so the second part of the $M_W$ matrix
is of a smaller size than the first. We find after some expansion work that
$$M_W=h\Bigl[\mtrix{\gamma_Ky(s)\alpha(s)/\theta_0^2 & 0 \cr 0 & 0 \cr}
+\delta_Kh^2\mtrix{b_{11} & b_{12} \cr b_{12} & b_{22} \cr}\Bigr]
+\mtrix{O(h^4) & O(h^5) \cr O(h^5) & O(h^5) \cr},$$
where $\gamma_K=\int K(u)^2\,{\d u}$ and $\delta_K=\int u^2K(u)^2\,{\d u}$,
and where $b_{11}=\half(y\alpha)''(s)/\theta_0^2$,
$b_{12}=(y\alpha)'(s)/\theta_0$, and $b_{22}=y(s)\alpha(s)$. 

To reach expressions for $J_W^{-1}M_WJ_W^{-1}$ we need to work with
a matrix of the form 
$(cE_{11}+h^2A)^{-1}(dE_{11}+h^2B)(cE_{11}+h^2A)^{-1}$, 
where $E_{11}$ is the matrix with 1 as (1,1) element and zeros elsewhere.
The result, after lengthy but elementary calculations, is of the form 
$$J_W^{-1}M_WJ_W^{-1}=\mtrix{h^{-1}\gamma_K\alpha(s)/y(s)+c_{11}h+O(h^2) 
& h^{-1}c_{12}+O(h) \cr
h^{-1}c_{12}+O(h) & h^{-3}(\delta_K/\beta_K^2)/\{y(s)\alpha(s)\}
	+h^{-1}c_{22}+O(1) \cr},$$
for certain $c_{ij}$. We are primarily interested in the approximate
variance for the local $\hatt\theta(s)$, in view of (4.4), and this is 
$(nh)^{-1}\gamma_K\alpha(s)/y(s)+O(h/n)$, precisely as in the
one-dimensional case (3.10). 
Hence bias and variance properties are of the same form as
in the one-dimensional case, but with a different bias factor,
inherited from the model one smooths. 
 
\subsection
{\csc 4.2. Dynamic likelihood for a general multi-parameter model.}
Suppose the hazard rate model is of the type $\alpha(t)=a\gamma(t,\beta)$,
i.e.~a constant parameter $a$ times a function which depends on 
a possibly multi-dimensional parameter $\beta$ but not on $a$. 
Reparametrise locally to 
$$\alpha(t)=a\gamma(s,\beta)\{\gamma(t,\beta)/\gamma(s,\beta)\}
 =\theta\exp\{C(t,\beta)-C(s,\beta)\}
\quad {\rm for\ }t\in[s-\half h,s+\half h]. \eqno(4.6)$$
The score type function of the model is 
$\psi(t,\theta,\beta)=(1/\theta,C^*(t,\beta)-C^*(s,\beta))$,
where $C^*(t,\beta)={\dell\over \dell\beta}C(t,\beta)$. 
Notice that the local estimate $\hatt\beta(s)$ is only 
`silently present', in that it is used only in conjunction 
with finding the local $\hatt\theta(s)$, as with (4.3) and (4.4). 

Now the calculations of the Gompertz model above can be repeated
with the necessary modifications. As in that case one finds
$\theta_0=\alpha(s)+\half\beta_Kh^2b(s,\beta_0)$ with a similar
$b(s,\beta_0)$, and also that $\theta_0c(s,\beta_0)=\alpha'(s)+O(h^2)$,
where $c(t,\beta)={\dell\over \dell t}C(t,\beta)$. This leads to 
$$\eqalign{
\E\hatt\alpha(s)&=\alpha(s)+\half\beta_Kh^2
\bigl[\alpha''(s)-\alpha(s)\{c(s,\beta_0)^2
+c'(s,\beta_0)\} \cr 
& \qquad +2\{y'(s)/y(s)\}\{\alpha'(s)-\alpha(s)c(s,\beta_0)\}\bigr]
+O(h^4) \cr
&=\alpha(s)+\half\beta_Kh^2
\bigl[\alpha''(s)-\theta_0\{c(s,\beta_0)^2+c'(s,\beta_0)\}
+2\{y'(s)/y(s)\}O(h^2)\bigr]+O(h^4) \cr
&=\alpha(s)+\half\beta_Kh^2\{\alpha''(s)-\alpha_0''(s)\}+O(h^4), \cr}$$
where $\alpha_0''(s)$ is the second derivative of the model's hazard rate
$\theta\exp\{C(t,\beta)-C(s,\beta)\}$ w.r.t.~$t$, 
evaluated at $s$, and with the local least false parameters
$\theta_0=\theta_0(W)$ and $\beta_0=\beta_0(W)$ inserted.  
We also have $\alpha_0''(s)=\theta_0\{c(s,\beta_0)^2+c'(s,\beta_0)\}
=\alpha'(s)^2/\alpha(s)+\alpha(s)c'(s,\beta_0)+O(h^2)$. 
Note that (4.5) is a special case. 

One next finds that the (1,1) element
of the appropriate $J_W^{-1}M_WJ_W^{-1}$ matrix is yet again
equal to $h^{-1}\gamma_K\alpha(s)/y(s)+O(h)$, albeit with a 
more involved expression for the constant in the secondary $O(h)$ term. 
The basic properties for the dynamic likelihood estimator 
are accordingly once more of the familiar type 
$$\E\hatt\alpha(s)=\alpha(s)
+\half\beta_Kh^2\{\alpha''(s)-\alpha_0''(s)\}+O(h^4)
\quad {\rm and} \quad 
\Var\,\hatt\alpha(s)\doteq
	{\gamma_K\over nh}{\alpha(s)\over y(s)}, \eqno(4.7)$$
with a bias term $\half\beta_Kh^2b(s)$ appropriate to the parametric
model employed. As noted above there are also alternative useful 
expressions for the $b(s)$ term, since we can move out $O(h^2)$ terms. 

\subsection
{\csc 4.3. A running Weibull estimator.}
As an example of the previous general machinery, consider the 
Weibull model, which uses $\alpha(t)=abt^{b-1}$ for certain parameters 
$a$ and $b$. We reparametrise to $\alpha(t)=\theta(t/s)^\beta$,
where $\theta=abs^{b-1}$ and $\beta=b-1$. Let $\hatt\theta(s)$ 
and $\hatt\beta(s)$ maximise the kernel smoothed dynamic log-likelihood
$$\int_WK(h^{-1}(t-s))\bigl[\{\log\theta+\beta(\log t-\log s)\}\,{\d N}(t)
	-Y(t)\theta(t/s)^{\beta}\,{\d t}\bigr].$$
Then use $\hatt\alpha(s)=\hatt\theta(s)$ in the end. 
The results above imply $\theta_0\beta_0=s\alpha'(s)+O(h^2)$, and 
the bias is 
$$\half\beta_Kh^2\{\alpha''(s)-\theta_0(\beta_0^2-\beta_0)/s^2\}+O(h^4)
 =\half\beta_Kh^2\{\alpha''(s)-\alpha'(s)^2/\alpha(s)+\alpha'(s)/s\}
 +O(h^4). \eqno(4.8)$$
The approximate variance is yet again $\gamma_K(nh)^{-1}\alpha(s)/y(s)$. 
Note that the bias is only $O(h^4)$ if the true hazard is locally 
a Weibull hazard. 
	
\subsection
{\csc 4.4. A dynamic nonincreasing estimator.}
As a final example, consider the simple frailty model with
hazard rate $a/(1+\beta t)$. This is the hazard rate in a population
where each individual has a constant hazard rate but where these
vary in the population according to a gamma distribution with 
mean $a$ and variance $\beta$. The local parametrisation is
$\theta(1+\beta s)/(1+\beta t)$ for $t\in[s-\half h,s+\half h]$. 
Even though the model can tolerate a small negative value for 
$\beta$ we shall in this example take it a priori as a nonnegative quantity. 	
So let $\hatt\theta(s)$ and $\hatt\beta(s)$ maximise 
$$\int_WK(h^{-1}(t-s))\bigl[\{\log\theta+\log(1+\beta s)-\log(1+\beta t)\}
\,{\d N}(t)-Y(t)(1+\beta s)\,{\d t}/(1+\beta t)\bigr], $$
and use $\hatt\alpha(s)=\hatt\theta(s)$ in the end. Then
$$\E\hatt\alpha(s)=\alpha(s)+\half\beta_Kh^2\{\alpha''(s)
	-2\alpha'(s)^2/\alpha(s)\}+O(h^4) 	
	\quad {\rm and} \quad
	\Var\,\hatt\alpha(s)\doteq{\gamma_K\over nh}{\alpha(s)\over y(s)}.
	\eqno(4.9)$$
	
\bigskip
{\bf 5. Comparison with the traditional kernel estimator.} 
Estimators developed in Sections 3 and 4 can now be compared with 
the classical nonparametric estimator.

\subsection
{\csc 5.1. The smoothed Nelson--Aalen estimator.}
The traditional nonparametric estimator is a kernel smooth of 
the (2.1) estimator of the cumulative, 
$$\tilda\alpha(s)=\int_Wh^{-1}K(h^{-1}(t-s))\,\d\hatt A(t)
=\sum_{|x_i-s|\le h/2}h^{-1}K(h^{-1}(x_i-s))\delta_i/Y(s_i). \eqno(5.1)$$
When $K(u)$ is uniform this becomes 
$\{\hatt A(s+\half h)-\hatt A(s-\half h)\}/h$, for example.  
From the properties of $\hatt A$ reviewed in Section 2 
it is not difficult to derive 
$$\E\tilda\alpha(s)\doteq\alpha(s)+\half\beta_Kh^2\alpha''(s)
	\quad {\rm and} \quad 
  \Var\,\tilda\alpha(s)\doteq 
	{\gamma_K\over nh}{\alpha(s)\over y(s)}. \eqno(5.2)$$
See also 
Ramlau-Hansen (1983), 
Yandell (1983), and 
Tanner and Wong (1983), who all studied estimators of this type, 
and Andersen et al.~(1993, Chapter IV). 
It is remarkable that the new estimators $\alpha(s,\hatt\theta(s))$ 
and the traditional one have exactly the same approximate variance 
and the same type of approximate bias,
when they use the same kernel and the same bandwidth;
see (3.10), (4.4) and (4.7).

\subsection
{\csc 5.2. When is the dynamic method always better?}
The dynamic kernel smoothed likelihood estimator has approximate 
bias $\half\beta_Kh^2b(s)$, with a $b(s)$ function depending 
on the underlying parametric family used. In view of the 
comparison already made above it follows that the new method
is always as good as or better than the 
Ramlau-Hansen--Yandell estimator, 
with the same kernel and the same window size, provided only 
$|b(s)|\le|\alpha''(s)|$. 

For the one-parameter situation the question is whether 
$$|\alpha''(s)-\alpha_0''(s)+2\{\alpha'(s)-\alpha_0'(s)\}
	\{y'(s)/y(s)+\psi_0'(s)/\psi_0(s)\}|\le|\alpha''(s)|. \eqno(5.3)$$
This can easily happen if the parametric family is only moderately acceptable. 
For the special case (3.12), 
for which $\alpha_0'(s)$, $\alpha_0''(s)$ and $\psi_0'(s)$ 
are absent, the inequality might take place 
in regions where $\alpha$ is convex and increasing, or concave and decreasing. 
When there is no censoring $y=\exp(-A)$ and 
$y'/y=-\alpha$, and then the criterion for when (3.12) is better than
then the traditional (5.1) becomes $0\le\alpha(s)\alpha'(s)/\alpha''(s)\le1$. 

For the multi-parametric families of Section 4 we have established 
$b(s)=\alpha''(s)-\alpha_0''(s)$, with $\alpha_0''(s)$ stemming from
the model used, and with certain useful alternative expressions. 
The dynamic likelihood estimator is better than (5.1), when the same $h$
is used, whenever $|\alpha''(s)-\alpha_0''(s)|\le|\alpha''(s)|$,
which can be rewritten 
$$0\le{\alpha_0''(s)\over \alpha''(s)}\le2 \eqno(5.4)$$ 
when the second derivative of the true hazard is not zero. 
If the parametric model used is locally correct, then the ratio is 1 
and the bias is $O(h^4)$ only. If we take `the parametric model
is roughly adequate' to mean (5.4), then indeed the dynamic likelihood
estimator is always better than (5.1) under such circumstances, 
for each $h$ and each $K$. 

For the two-parametric running Gompertz estimator (4.4) the criterion is
$$0\le\alpha'(s)^2/\{\alpha(s)\alpha''(s)\}\le2, \eqno(5.5)$$
and the ratio is 1 exactly for Gompertz hazards $a\exp(\beta s)$.
If for example $\alpha(s)=a+be^{cs}$ is of Gompertz--Makeham form,
then the ratio is $be^{cs}/(a+be^{cs})$ and well inside $(0,2)$,
showing that (4.4) will be better than (5.1) for all such hazards.
Similarly, for the two-parametric running Weibull estimator of 4.3,
the new estimator is always better than (5.1) in regions where 
$$0\le \alpha'(s)^2/\{\alpha(s)\alpha''(s)\}
	-\alpha'(s)/\{s\alpha''(s)\}\le2, \eqno(5.6)$$
and the function appearing in the middle is equal to 1 
exactly for Weibull hazards. And finally the criterion for when
the estimator of 4.4 is always better than (5.1) is 
$$0\le\alpha'(s)^2/\{\alpha(s)\alpha''(s)\}\le 1. \eqno(5.7)$$

\subsection
{\csc 5.3. Vicinity of parametric model.}
As explained above one can expect the methods developed to perform 
better than the traditional (5.1), and surely also better than
other purely nonparametric estimators, 
if only the parametric model used is roughly adequate. 
So far this statement has been referring to a comparison 
when the two methods have used the same window width $h$. 
But in these cases we would really expect the new methods to 
perform not only better but much better, by carefully choosing 
a good window width. When the bias is smaller we can select 
a larger window and be rewarded with smaller variability,
cf.~the mean squared error calculations of 6.1. 

It should also be possible to improve on the convergence rate 
if the true hazard lies suitably close to the parametric family. 
A mathematical framework to make this notion more precise
could be as follows. There is a sequence of experiments 
where at stage $n$ there are data $(X_i,\delta_i)$ on $n$ individuals 
coming from a distribution with true hazard $\alpha(.)=\alpha_n(.)$. 
Suppose this is such that the bias factor is $b(s)=b_0(s)n^{-\epsilon}$,
for some $\epsilon\in[0,\half)$. Then the best achievable 
mean squared error is of size proportional to $n^{-(4-2\epsilon)/5}$,
and this happens with $h$ chosen as a suitable $h_0=cY_n(s)^{-(1-2\epsilon)/5}$.
A cross validation or other clever $h$ selection scheme will
pick this up, cf.~the following section. The best nonparametric
convergence rate is $n^{-4/5}$, for both point-wise and integrated 
mean squared error, and these calculations show that this
can be improved upon for alternatives in the vicinity of the parametric
model. If $\alpha''(s)-\alpha_0''(s)=O(n^{-1/4})$, for example,
then the mean squared error is $O(n^{-9/10})$, and alternatives
lying almost $O(n^{-1/2})$ away, in the above sense, are estimated 
with almost full parametric $O(n^{-1})$ precision. The point of comparison
is that the traditional (5.1) estimator will still only accomplishes
$O(n^{-4/5})$ precision for these hazard rates. 
			
\bigskip
{\bf 6. Choosing the smoothing parameter.}
We have defined $\hatt\alpha(s)=\alpha(s,\hatt\theta(s))$ for 
given parametric family $\alpha(s,\theta)$ and kernel $K$.
The most decisive influence on the estimator is due to
the smoothing parameter $h$. 

\subsection
{\csc 6.1. Mean squared error calculations.} 
By (3.10) and (4.7) the approximate mean squared error is of the form 
$$\E\{\hatt\alpha(s)-\alpha(s)\}^2
	\doteq{1\over4}h^4\beta_K^2b(s)^2
	+{\gamma_K\over nh}{\alpha(s)\over y(s)}, $$ 
where $b(s)$ is the appropriate bias factor stemming from the parametric
recipe used. The mean squared error is minimised for 
$$h_0(s)=\Bigl\{{\gamma_K\over \beta_K^2}{\alpha(s)\over b(s)^2}\Bigr\}^{1/5}
	{1\over \{ny(s)\}^{1/5}}. \eqno(6.1)$$
The resulting minimal mean squared error is 
${5\over4}(\beta_K\gamma_K^2)^{2/5}\alpha(s)^{4/5}b(s)^{2/5}
	/\{n\allowbreak y(s)\}^{4/5}$.
Different choices of reasonable kernels give about the 
same result, but the best choice, managing to minimise $\beta_K\gamma_K^2$
among kernels on $[-\half,\half]$ with integral 1, 
is the Bartlett--Yepanechnikov kernel 
$K_0(u)={3\over2}(1-4u^2)$ on $[-\half,\half]$. 
The resulting $\alpha(s,\hatt\theta(s))$ estimator 
is continuous in $s$ but its derivative will have discontinuities 
at points $s$ where $s\pm\half h$ hits an observed failure time. 

We have seen that the new methods can outperform the traditional ones
by reducing the bias, say from $b_{\rm trad}(s)=\alpha''(s)$ of (5.1)
to possibly smaller $b(s)=\alpha''(s)-\alpha_0''(s)$ for those of Section 4.
It is therefore of interest to note that the squared bias makes up
20\% and the variance 80\% of the approximate mean squared error, 
so bias reduction can perhaps not be expected to give dramatic gains. 
If $b(s)=\half b_{\rm trad}(s)$, for example, then the best theoretical
window width becomes $h_0=1.32\,h_{0,{\rm trad}}$, and the 
mean squared error is reduced with 24\%. 

The $h_0$ formula cannot be put to direct use since it depends on the 
hazard rate itself, but the rate of convergence to zero of mean
squared error becomes the optimal $n^{-4/5}$ when $h$ is chosen 
proportional to $n^{-1/5}$. 
The formula indicates that $h$ should be chosen 
proportional to $Y(s)^{-1/5}$ in practice. 
One possibility is to use $h_n=cY(s)^{-1/5}$ and try to 
minimise a global criterion like 
$\E\int_0^Tw(s)\{\hatt\alpha(s)-\alpha(s)\}^2\,{\d s}$ w.r.t.~$c$. 
The result is a local variable kernel smoothed likelihood estimator with 
$$h_n=\Bigl\{{\gamma_K\over \beta_K^2}
	{\int_0^Tw(s)y(s)^{-4/5}\alpha(s)\,{\d s} \over 
	\int_0^T w(s)y(s)^{-4/5}b(s)^2\,{\d s}}\Bigr\}^{1/5}
	{1\over Y(s)^{1/5}}. \eqno(6.2)$$
We might for example choose weight function $w(s)=y(s)^{4/5}$ here,
this being inversely proportional to the optimal mean squared error,
and this simplifies (6.2).  
The nominator integral can be estimated with $n^{1/2}$-precision. 
Some pilot estimate $\hatt\alpha_{\rm pil}$, like the (5.1) estimator
with an overall twice differentiable kernel $K_2$ and a somewhat large $h_2$, 
can be used to estimate the denominator integral. 
A final adjustment is needed since $\int_0^T\hatt b(s)^2\,{\d s}$ 
will be biased. 
Working out expressions for the bias of $\int_0^T\hatt b(s)^2\,\d s$
as an estimator of $\int_0^Tb(s)^2\,\d s$ takes some efforts, 
for the most interesting estimators of Sections 3 and 4, 
but is within comfortable reach of Ramlau-Hansen's (1983) methods. 
In the end this produces a practical algorithm of `plug-in' type.  


The discussion above is valid for one-parameter families and also for
the class of multi-parameter families considered in Section 4,
since the approximate variance of $\hatt\alpha(s)$ 
also in these situations turned out to be of 
the form $\gamma_K(nh)^{-1}\alpha(s)/y(s)$. 
In the models of Section 4 there is a `local position' parameter 
$\theta$ and a `local slope' parameter $\beta$. 
Note that the local slope estimate $\hatt\beta(s)$ 
has quite larger variance than the local position estimate $\hatt\theta(s)$.
In the running Gompertz case the slope estimate has variance 
proportional to $n^{-1}h^{-3}/\{y(s)\alpha(s)\}$, for example. 
The best window size for $\beta$ estimation
is proportional to $Y(s)^{-1/7}$, but the best size for 
$\theta$ estimation, which is our primary concern, is still
proportional to $Y(s)^{-1/5}$. These quantitative results are 
perhaps as expected, in view of similar results from density 
estimation and nonparametric regression. They also suggest that 
the $\hatt\beta(s)$ that is inserted in $\hatt\theta(s,\beta)$ of (4.3) 
to produce the final (4.4) can be quite variable if produced from
$Y(s)^{-1/5}$-windows, and it may be advantageous to use a separate
estimation scheme for estimation of this parameter, with somewhat
larger windows. See also Remarks 7B and 7G.

The reasoning that led to (6.1) and (6.2) is also pertinent for 
the problem of choosing $h$ in the Ramlau-Hansen--Yandell estimator (5.1),
since the bias and variance structure are of the same type, 
only with $b(s)=\alpha''(s)$ instead. 
We also note that there are other ways of obtaining a data-driven $h_n(s)$,
like cross validation or bootstrapping, 
but these are not discussed further here. References to cross validation
techniques for the (5.1) estimator are Nielsen (1990) and Gr\'egoire (1993), 
and these techniques should carry over at least to the (3.12) estimator.

\subsection
{\csc 6.2. Local goodness of fit testing.}
If the parametric model doesn't fit well the dynamic likelihood
hazard estimator is still reasonable, and resembles the 
nonparametric Nelson--Aalen smoother (5.1) in performance. 
At the same time our method is able to outperform 
(5.1) as well as other purely nonparametric methods
in cases where the parametric family 
$\alpha(s,\theta)$ used is only roughly acceptable, 
as explained in Section 5. In such cases the size of the bias is small, 
which by (6.2) suggests using quite a large bandwidth $h$, 
which in its turn almost amounts to using an ordinary parametric method. 
 
A natural but somewhat elaborate strategy 
is to choose $h=\hatt h(s)$ to be the smallest $h$ for which 
some convenient goodness of fit criterion rejects the 
parametric model on $s\pm\half h$. The ultimate case is of course
no detectable departure from the model over the full range $[0,T]$, 
which then leads to using $h=\infty$, i.e.~ordinary parametric
estimation $\alpha(s,\hatt\theta_{[0,T]})$, say. 

Hjort (1985, 1990) has developed classes of goodness of fit tests
for general parametric counting process models, and these are indeed
presented there as tests of validity over the full range $[0,T]$. 
Similar mathematical techniques can however be used to construct
procedures that check model adequacy 
over a general $[a,b]$ interval, and some such are presented next. 
This apparatus would then be used with $[a,b]=[s-\half h,s+\half h]$, mostly,
but to get the running estimator started one would look for
model adequacy over $[0,b]$ intervals first, cf.~Remark 7A. 

\subsection
{\csc 6.3. One-parameter families.}
Consider dynamic smoothing of an arbitrary 
one-dimensio\-nal parametric family $\alpha(u,\theta)$. 
Let $\hatt\theta_{[a,b]}$ be the local maximum likelihood estimator using 
only $[a,b]$ information, i.e.~it solves 
$\int_a^b\psi(u,\theta)\{{\d N}(u)
-Y(u)\alpha(u,\theta)\allowbreak\,{\d u}\}=0$. 
Let 
$$D_n(t)=n^{-1/2}\int_a^t\psi(u,\hatt\theta_{[a,b]})\{{\d N}(u)
	-Y(u)\alpha(u,\hatt\theta_{[a,b]})\,{\d u}\} 
	\quad {\rm for\ }t\in[a,b].$$
It uses the `basic martingale' $\d N(u)-\alpha(u,\theta)\,\d u$
and is able to pick up departures from the parametric model. 
Notice that $D_n(.)$ starts and ends at zero. 
Methods of Hjort (1990) can be used to prove that $D_n(.)$,
if indeed the model holds on $[a,b]$, converges to a zero-mean
Gau\ss ian process $D(.)$ with covariance function 
${\rm cov}\{D(t_1),D(t_2)\}=\tau^2(b)\{p(t_1\wedge t_2)-p(t_1)p(t_2)\}$,
in which 
$\tau^2(t)=\int_a^ty(u)\psi(u,\theta)^2\alpha(u,\theta)\allowbreak\,{\d u}$
and $p(t)=\tau^2(t)/\tau^2(b)$. But this shows that $D(.)$ 
is distributed as a scaled and time-transformed Brownian bridge,
$\tau(b)W^0(p(.))$. Consequently $\max_{a\le t\le b}|D_n(t)|/\hatt\tau(b)$
is asymptotically distributed as $\|W^0\|=\max_{0\le s\le 1}|W^0(s)|$,
where 
$\hatt\tau^2(b)=\int_a^bn^{-1}Y(u)\psi(u,\hatt\theta_{[a,b]})^2
\allowbreak\alpha(u,\hatt\theta_{[a,b]})\,{\d u}$ estimates $\tau^2(b)$. 
A natural procedure is therefore to stretch the 
$[a,b]=[s-\half h,s+\half h]$ interval until 
$$\Bigl\{\int_a^bY\!\!(u)\psi(u,\hatt\theta_{[a,b]})^2
\alpha(u,\hatt\theta_{[a,b]})\,{\d u}\Bigr\}^{-1/2}\max_{a\le t\le b}
\Big|\int_a^t\!\!\psi(u,\hatt\theta_{[a,b]})\{{\d N}(u)
-Y(u)\alpha(u,\hatt\theta_{[a,b]})\,{\d u}\}\Big|
\ge 1.225, \eqno(6.3)$$
say, 1.225 being the upper 10\% point of the distribution of $\|W^0\|$. 
One might opt for 1.359 instead, the upper 5\% point. 
Observe that the maximum value must be attained at one of 
the points $x_i$ of $x_i-$, with $a\le x_i\le b$, so the 
continuous maximum is really only a finite maximum, and is 
perfectly feasible to compute efficiently, for given $s\pm\half h$ window.

When choosing window sizes for the (3.12) estimator, for example,
which uses local constants, the windows should be stretched until
$$N[a,b]^{-1/2}\max_{a\le t\le b}\Big|N[a,t]
	-\int_a^tY(u)\hatt\theta_{[a,b]}\,{\d u}\Big|\ge 1.225, \eqno(6.4)$$
where in this case $\hatt\theta_{[a,b]}=N[a,b]/\int_a^bY(u)\,{\d u}$. 

\subsection
{\csc 6.4. Local model adequacy for multi-parameter hazard rates.}
Next turn attention to dynamic likelihood smoothing of a 
multi-parametric class of hazards, with $p\ge2$ parameters. 
Let this time
$$D_n(t)=n^{-1/2}\Bigl\{N[a,t]-\int_a^tY(u)\alpha(u,\hatt\theta_{[a,b]})
	\,{\d u}\Bigr\} \quad {\rm for\ }t\in[a,b], $$
with a view towards using the maximal absolute value as a test
for model adequacy. Here $\hatt\theta_{[a,b]}$ is the 
local maximum likelihood estimator using $[a,b]$-information, 
i.e.~solving the $p$ equations 
$\int_a^b\psi(u,\theta)\{{\d N}(u)-Y(u)\alpha(u,\theta)\,{\d u}\}=0$.
Techniques of Hjort (1990) can be used 
to demonstrate process convergence of $D_n(.)$ towards
$$D(t)=V[a,t]-\Bigl(\int_a^ty(u)\psi(u,\theta)\alpha(u,\theta)\,{\d u}\Bigr)'
	\Sigma^{-1}\int_a^b\psi(u,\theta)\,{\d V}(s), $$
where $V(.)$ is a Gau\ss ian martingale with noise level
$\Var\,{\d V}(u)=y(u)\alpha(u,\theta)\,{\d u}$, and where 
$\Sigma=\int_a^by(u)\psi(u,\theta)\psi(u,\theta)'\alpha(u,\theta)\,{\d u}$. 
The $D_0(t)=V[a,t]=\int_a^t{\d V}(u)$ process is quite simple, it has
independent increments and hence is a scale- and time-transformed 
Brownian motion process. The point is now that if one considers
the $D_0(.)$ process conditioned on the $p$ events 
$\int_a^b\psi(u,\theta)\,{\d V}(u)=0$, then Gau\ss ianeity and covariance
calculations can be furnished to demonstrate that 
this is exactly distributed as the $D(.)$ process; 
cf.~Remark 7F in Hjort (1990). 
This makes it possible to bound the distribution
of $\|D\|=\max_{a\le t\le b}|D(t)|$, even though the exact distribution 
might be too difficult to obtain. 
	
We now specialise to a class of hazards of the form 
$\alpha(u,\theta)=\theta\gamma(u,\beta)$, cf.~(4.6),
in which case the $\psi(.)$ function has first component $1/\theta$
and second component $\phi(u,\beta)$, say.
In this case the limit process $D(.)$ is distributed as $D_0(.)$,
tied down first with $D_0(b)=0$ and then with 
$\int_a^b\phi(u,\beta)\,{\d V}(u)=0$. Letting $D^*(.)$ be the result
of tying down $D_0(.)$ with only the first requirement, covariance
calculations show that $D^*(.)=_d\tau(b)W^0(p(.))$, 
this time with $\tau^2(t)=\int_a^ty(u)\theta\gamma(u,\beta)\,{\d u}$ and
$p(t)=\tau^2(t)/\tau^2(b)$. So the distribution of $D(.)$ 
is that of tying down $D^*(.)$ further, and it can be seen that
the distribution of $\|D\|$ is stochastically smaller than
the distribution of $\|D^*\|$, just as the distribution of
a maximal absolute Brownian bridge is stochastically smaller than
the distribution of a maximal absolute Brownian motion. 
Here $\tau^2(b)$ is estimated consistently with
$\int_a^bn^{-1}Y(u)\hatt\theta_{[a,b]}\gamma(u,\hatt\beta_{[a,b]})\,{\d u}$,
and we also have $\hatt\theta_{[a,b]}=N[a,b]/\int_a^bY(u)\gamma(u,
\hatt\beta_{[a,b]})\,{\d u}$. The end result is to use
$$N[a,b]^{-1/2}\max_{a\le t\le b}\Big|N[a,t]
	-\int_a^tY(u)\hatt\theta_{[a,b]}\gamma(u,\hatt\beta_{[a,b]})
	\,{\d u}\Big|\ge 1.225 \eqno(6.5)$$
as a conservative 10\% level test criterion for rejecting 
$\theta\gamma(u,\beta)$ as a model for the hazard on $[a,b]$. 

It is worth noting that the difference
between the distributions of $\|D\|$ and $\|D^*\|$ is small when
the $[a,b]$ interval is not large, provided the model being
tested has the local reparametrisation form (4.6). 
This can be shown after expanding the $\Sigma^{-1}$ matrix here in a way 
similar to that for $J_W^{-1}$ in Sections 4.1 and 4.2. 
Thus 1.225 above is meant to be a conservative value but actually
also an approximation to the real 0.90 point of the null distribution.
Of course this approximation cannot be expected to be overly precise,
and some experimentation with the 1.225 rejection limit would be needed. 
On the computational side we point out that the maximum again must be 
attained for one of $t=x_i$ or $x_i-$ with $a\le x_i\le b$. Furthermore,
$$\int_a^{x_i}Y(u)\hatt\theta_{[a,b]}\gamma(u,\hatt\beta_{[a,b]})\,{\d u}
=\sum_{j\colon x_j\ge a}\hatt\theta_{[a,b]}
\{G(x_i\wedge x_j,\hatt\beta_{[a,b]})-G(a,\hatt\beta_{[a,b]})\}, \eqno(6.6)$$
where $G(t,\beta)=\int_0^t\gamma(u,\beta)\,{\d u}$. 

\subsection
{\csc 6.5. Other tests for model adequacy on intervals.} 
There are naturally other possible goodness of fit tests for intervals,
see Hjort (1990) for other $D_n(.)$ type functions 
and for classes of chi squared type tests and 
Hjort and Lumley (1993) for normalised local hazard plots. 
Chi squared methods would be awkward to implement in a general
way here, since the $[a,b]$ intervals would often be short. 
We record a couple of potentially useful variations on the 
$D_n(.)$ theme, however, with a view towards quick calculations
and decisions, since tests are to be carried out on slowly
expanding $[s-\half h,s+\half h]$ intervals for each $s$. 

(6.3)--(6.5) arose as maxima of the $D_n(.)$ process, 
and utilised convergence to suitably scaled and time-transformed 
Brownian bridges, as with Kolmogorov--Smirnov type tests. 
Martingale techniques for the counting process $N$ can be used to show 
$$\eqalign{
\int_a^b|D_n(t)|^q\,{\d N}(t)/n
&\arr_d\int_a^b|D(t)|^qy(t)\theta\gamma(t,\beta)\,{\d t} \cr
&\le_d\int_a^b|\tau(b)W^0(p(t))|^q\tau^2(b)\,dp(t)
 =\tau(b)^{2+q}\int_0^1|W^0(s)|^q\,{\d s}, \cr} $$
where `$\le_d$' means `stochastically smaller than'. 
For $q=2$ we have a Cram\'er--von Mises type test, with rejection criterion
$$\eqalign{
\hatt\tau(b)^{-4}\int_a^bD_n(t)^2\,{\d N}(t)/n
&={n^{-1}\sum_{a\le x_i\le b}D_n(x_i)^2\delta_i
\over (n^{-1}N[a,b])^2} \cr
&={\sum_{a\le x_i\le b}\{N[a,x_i]-\int_a^{x_i}Y(u)\hatt\theta_{[a,b]}
\gamma(u,\hatt\beta_{[a,b]})\,{\d u}\}^2\delta_i
\over N[a,b]^2}\ge 0.347 \cr}\eqno(6.7)$$
on the 10\% significance level. With wished for 5\% level we would use
0.461 instead, the 0.95 quantile of the $\int_0^1W^0(s)^2\,{\d s}$ distribution.
The second variation is for $q=1$, where we use 
$$\eqalign{
\hatt\tau(b)^{-3}\int_a^b|D_n(t)|\,{\d N}(t)/n
&={n^{-1}\sum_{a\le x_i\le b}|D_n(x_i)|\delta_i
\over (n^{-1}N[a,b])^{3/2}} \cr
&={\sum_{a\le x_i\le b}\big|N[a,x_i]-\int_a^{x_i}Y(u)\hatt\theta_{[a,b]}
\gamma(u,\hatt\beta_{[a,b]})\,{\d u}\big|
\over N[a,b]^{3/2}}\ge 0.499 \cr}\eqno(6.8)$$
for intended 10\% significance level, and 0.582 for intended 5\% significance
level. 0.499 and 0.582 are upper quantiles of the $\int_0^1|W^0(s)|\,{\d s}$
distribution. 
There are simpler one-parameter analogues to (6.7) and (6.8),
essentially as in these formulae but with $\gamma(u,\beta)=1$.
Note that (6.6) can be used when computing any of these test statistics.  

Some experimentation with these $\hatt h=\hatt h(s)$ selectors 
is necessary. One should avoid using too small windows since 
this would lead to too irregular local estimates. 
We should therefore only search for acceptable windows 
$s\pm\half h$ with $h$ at least as large
as some suitably determined $h_0(s)$. One possibility is to demand
at least $k$ observed $x_i$'s in the window, say with $k=10$. 
Hence the (6.3)--(6.5) and (6.7)--(6.8) 
stopping criteria are to be used with such a modification. 
Secondly the realised $\hatt h(s)$ could be somewhat irregular 
as a function of $s$. A natural modification is to smooth 
this curve first, before finally computing the local likelihood
estimate $\alpha(s,\hatt\theta_{[s-\hatt h(s)/2,s+\hatt h(s)/2]})$. 

\bigskip
{\bf 7. Supplementing remarks.}

\medskip
{\csc 7A. Starting the estimator.}
We have defined $\hatt\alpha(s)=\alpha(s,\hatt\theta(s))$ with
parameter estimate obtained from $s\pm\half h$ data, which also
means that a separate definition is required for $s\le \half h$. 
One natural strategy is to use the model adequacy on intervals
methods of Sections 6.3--6.5 to find the smallest $b$ for which 
the model is rejected on $[0,b]$, and then use 
$\hatt\alpha(s)=\alpha(s,\hatt\theta_{[0,b_0]})$ for $s\in[0,b_0]$,
with a somewhat smaller $b_0$ than $b$. Another possibility is 
to use $\alpha(s,\hatt\theta(\half h))$ on $[0,\half h]$. 

\medskip
{\csc 7B. Post-smoothing of parameter estimates.}
The basic estimator is $\hatt\alpha(s)=\alpha(s,\hatt\theta(s))$ where
$\hatt\theta(s)$ uses only $s\pm\half h$ information. 
It is useful in practical applications to display not only the
final $\hatt\alpha(s)$ but also the parameter estimate function
or functions $\hatt\theta(s)$. Sometimes this function has discontinuities,
cf.~(3.12) and the requirements on $K$ noted there to give
smoothness. A general alternative is to post-smooth the 
parameter estimates, before plugging in to give $\hatt\alpha(s)$. 
Comments in 5.1, for example, suggest using post-smoothing of 
$\hatt\beta(s)$ in (4.3) and (4.4). 

\medskip
{\csc 7C. Density estimation with dynamic likelihood.}  
When $\alpha(.)$ is estimated one can of course also estimate
other quantities depending on $\alpha(.)$. 
The local likelihood methods of Sections 3 and 4 therefore apply to 
nonparametric or semiparametric density estimation as well, via the 
$f(t)=\alpha(t)\exp\{-A(t)\}$ connection. 
Methods given there can be used to obtain a locally estimated 
normal density of the type 
$\hatt f(t)=N\{\hatt\mu(t),\hatt\sigma(t)^2\}(t)$, for example.
There are at least two general immediate possibilities, namely
$$\eqalign{
\hatt f_1(t)&=\alpha(t,\hatt\theta(t))\exp\{-A(t,\hatt\theta(t)\}
\quad {\rm and} \cr 
\hatt f_2(t)&=\Bigl[\prod_{[0,t)}\{1-\alpha(s,\hatt\theta(s))\,{\d s}\}\Bigr]
	\alpha(t,\hatt\theta(t))
	=\exp\Bigl\{-\int_0^t\alpha(s,\hatt\theta(s))\,{\d s}\Bigr\}
	\alpha(t,\hatt\theta(t)). \cr}$$
The simplest case would again be that of a locally constant hazard,
for which 
$$\hatt f_1(t)=\hatt\theta(t)\exp\{-\hatt\theta(t)\,t\}
	\quad {\rm and} \quad 
\hatt f_2(t)=\exp\Bigl\{-\int_0^t\hatt\theta(s)\,{\d s}\Bigr\}
	\,\hatt\theta(t).$$
These are somewhat cumbersome density estimators. 
There are better schemes  
more directly geared towards the density estimation problem,
but still with the same local likelihood characteristics,
see Hjort and Jones (1993).  

\medskip
{\csc 7D. Regression models.}
Methods of this paper can be made to work in situations
with covariate information. Consider the Cox regression model
where individual $i$ has hazard rate of the form
$$\alpha_i(s)=\alpha_0(s)\exp(\beta'z_i)
\quad {\rm for\ }s\in[0,T]\ {\rm and}\ i=1,\ldots,n. $$
The $\alpha_0(.)$ function is the hazard rate for individuals with
covariate vector $z=0$, and is left unspecified. This baseline hazard
function can now be estimated using dynamic likelihood. If we fit 
a local constant on window $W=s\pm\half h$ the recipe is to 
maximise the kernel smoothed log-likelihood 
$$\sum_{i=1}^n\int_WK(h^{-1}(t-s))
\{(\log\theta+\beta'z_i)\,{\d N}_i(t)-Y_i(t)\theta\exp(\beta'z_i)\,{\d t}\}, $$
where ${\d N}_i(t)=I\{x_i\in[t,t+{\d t}],\delta_i=1\}$ and 
$Y_i(t)=I\{x_i\ge t\}$ are the 0--1 counting process and at risk process
for individual $i$. This gives 	
$$\hatt\alpha_0(s)={\sum_{i=1}^n\int_WK(h^{-1}(t-s))\,{\d N}_i(t)
\over \sum_{i=1}^n\int_WK(h^{-1}(t-s))Y_i(t)\exp(\hatt\beta'z_i)\,{\d t}}. $$
Here $\hatt\beta$ could be evaluated only locally, but if one
trusts the Cox model then $\beta$ remains constant over the
$[0,T]$ range, and we should accordingly use the same
$\hatt\beta$ regardless of $s$. But this is the same as smoothing
the traditional Breslow estimator. One can similarly construct 
a nonparametric $\alpha_0(.)$ estimator by fitting a running Weibull 
$\theta \gamma s^{\gamma-1}$, for example. The result is of the form 
$$\hatt\alpha_0(s)={\sum_{i=1}^n\int_WK(h^{-1}(t-s))\,{\d N}_i(t)
\over \sum_{i=1}^n\int_WK(h^{-1}(t-s))Y_i(t)\hatt\gamma(s) t^{\hatt\gamma(s)-1}
	\exp(\hatt\beta'z_i)\,{\d t}}. $$

Dynamic likelihood methods can also be developed in Aalen's
linear hazard rate regression model, by local parametric modelling
of the hazard factor functions. See Hjort (1993a). 

\medskip
{\csc 7E. Moderately incorrect parametric models.} 
A parametric model does not have to be fully perfect in order
for the methods based on it to be better than more conservative ones. 
In Hjort (1993b) a `tolerance distance' is calculated 
from a moderately incorrect model to a wider and
correct one; inside the tolerance radius 
estimators based on the incorrect model 
are better than those based on the correct model. 
For an example, suppose the true model is 
the gamma one, with hazard function inherited from the density 
$f(s,\theta,\gamma)=\{\theta^\gamma/\Gamma(\gamma)\}\,
	s^{\gamma-1}\exp(-\theta s)$. 
Then estimators based on the incorrect assumption of a constant rate 
(which corresponds to $\gamma=1$) 
are better than the two-parameter methods 
if $|\gamma-1|\le 1.245/\sqrt{n}$ (assuming no censoring). 
This can be seen as yet another argument for not giving up 
simple parametric methods, 
even though the underlying models might be wrong. 

\medskip
{\csc 7F. Counting process models.}
Methods and results of this paper can be generalised in various directions. 
They could be developed for Aalen's general multiplicative intensity
model for counting processes, and hence be used to estimate
hazard transition rates in time-inhomogeneous Markov chains, for 
example. There will then be a more complicated expression for the 
$M_W$ matrix of (2.6), but otherwise there will be few complications. 
In another direction our results could be extended to 
the full {half}line $[0,\infty)$ with appropriate 
extra assumptions on the censoring mechanism. 

\medskip
{\csc 7G. More theory.}
In our presentation we have concentrated on the perhaps most
immediate aspects of the dynamic likelihood estimation method. 
There are further natural questions to ask and further natural
results to prove. 
(i) One can prove uniform consistency of the 
(3.12) estimator without too much work, for example.
One can more generally establish 
$\max_{a\le s\le b}|\alpha(s,\hatt\theta(s))-\alpha(s)|\arr_p0$ 
under natural conditions. 
(ii) And the approximate size and distribution of this maximal
deviation quantity are also of interest. 
(iii) It is not difficult to establish that 
$(nh)^{1/2}\{\hatt\alpha(s)-\alpha(s)-\half\beta_K\alpha''(s)\}$
has a limiting zero-mean normal distribution, 
when $h\arr0$ and $nh\arr\infty$. 
This can also be used to construct point-wise approximate 
confidence band for the $\alpha(.)$ function, for example 
incorporating a bias correction $-\half\beta_K\hatt\alpha''(s)$.  
(iv) One should work out a reliable cross validation method for
minimising a nearly unbiased estimate of 
$\int_0^Tw(s)\{\hatt\alpha(s)-\alpha(s)\}^2\,\d s$,
say, as a function of the window width $h$, or as a function of 
$c$ in $h=cY_n(s)^{-1/5}$. The crux is to estimate 
$\int_0^Tw(s)\hatt\alpha(s)\alpha(s)\,\d s$. 
(v) Theory can also be worked out for estimation of derivatives
of the hazard function, as touched on in 6.1. Taking the derivative
of (3.12) to define $\hatt\alpha'(s)$, with a smooth kernel function $K$,
one can show that the bias is proportional to 
a $h^2b_1(s)$ and that the variance is proportional to 
$n^{-1}h^{-3}\alpha(s)/y(s)$, but with a $b_1(s)$ function different 
from that of the derivative of the Ramlau-Hansen--Yandell estimator. 
   
\medskip 
{\csc 7H. Questions.}
A simulation study comparing the various estimators would be welcome. 
Some of the questions to answer include:
How much better are the new estimators than the purely nonparametric ones 
when the true hazard is in the vicinity of the parametric model used?
How much do they lose to the parametric ones on the latter's home turf? 
Are there significant advantages to using multi-parameter models 
for the dynamic likelihood methods of Sections 3 and 4? 
What are the most useful ways of choosing window width $h=h_n(s)$?  

\bigskip
\bigskip
 
\parindent0pt
\parskip3pt
\baselineskip12pt

\def\ref#1{{\noindent\hangafter=1\hangindent=20pt
  #1\smallskip}}  
  
\centerline{\bf References}

\medskip


\ref{%
Andersen, P.K., Borgan, \O., Gill, R.D., and Keiding, N.L. (1993).
{\sl Statistical Models Based on Counting Processes.}
Springer-Verlag, New York.} 

\ref{%
Borgan, \O. (1984). Maximum likelihood estimation in parametric counting
process models, with applications to censored failure time data.
{\sl Scand.~J.~Statist.}~{\bf 11}, 1--16. Corrigendum, {\it ibid.}~p.~275.}



\ref{%
Gr\'egoire, C. (1993).
Least squares cross validation for counting process intensities.
{\sl Scand.~J.~Sta\-tist.}, to appear.} 

\ref{%
Hjort, N.L. (1985). 
Contribution to the discussion of Andersen and Borgan's 
`Counting process models for life history data: a review'.
{\sl Scand.~J.~Statist.}~{\bf 12}, 141--150.}







\ref{%
Hjort, N.L. (1990). 
Goodness of fit tests in models for life history data based on
cumulative hazard rates. 
{\sl Ann.~Statist.}~{\bf 18}, 1221--1258.} 


\ref{%
Hjort, N.L. (1991).
Semiparametric estimation of parametric hazard rates.
In {\sl Survival Analysis: State of the Art}, 
Kluwer, Dordrecht, pp.~211--236. 
Proceedings of the {\sl NATO Advanced Study Workshop on Survival
Analysis and Related Topics}, Columbus, Ohio, eds.~P.S. Goel and J.P.~Klein.} 

\ref{%
Hjort, N.L. (1992). 
On inference in parametric survival data models. 
{\sl Int.~Statist.~Review} {\bf 60}, 355--387.}

\ref{%
Hjort, N.L. (1993a).
Efficiency of three estimators in Aalen's linear
hazard rate regression model.
Submitted for publication.}

\ref{%
Hjort, N.L. (1993b).
Estimation in moderately misspecified models.
Submitted for publication.} 

\ref{%
Hjort, N.L.~and Jones, M.C. (1993). 
Locally parametric nonparametric density estimation.
Manus\-cript in progress.}

\ref{%
Hjort, N.L.~and Lumley, T. (1993).
Normalised local hazard plots.
Submitted for publication.} 


\ref{%
Nielsen, J.P. (1990).
Kernel estimation of densities and hazards: a counting process approach.
PhD.~dissertation, Department of Statistics,
University of Berkeley, California.}  


\ref{%
Ramlau-Hansen, H. (1983). 
Smoothing counting process intensities by means of kernel functions.
{\sl Ann.~Statist.}~{\bf 11}, 453--466.} 

\ref{%
Tanner, M.A.~and Wong, W.H. (1983).
The estimation of the hazard function from randomly censored data.
{\sl Ann.~Statist.}~{\bf 11}, 989--993.} 

\ref{%
Yandell, B.S. (1983).
Nonparametric inference for rates and densities with censored serial data.
{\sl Ann.~Statist.}~{\bf 11}, 1119--1135.} 

\bye